\documentclass[aip,twocolumn,letterpaper]{revtex4}

\usepackage{fullpage}

\usepackage{graphics}
\usepackage{epsfig}

\begin{document}
\title{Problematic nature of plasmoid theory of magnetic reconnection}
\author{Allen H. Boozer}
\affiliation{Columbia University, New York, NY  10027\\ ahb17@columbia.edu}

\begin{abstract}

Plasmoid theory uses two-coordinate models to explain fast magnetic reconnection, which occurs on an Alfv\'enic, not a resistive time scale, in plasmas that are evolving in the near-ideal limit.   The primary application has been to three-dimensional naturally-occurring  plasmas.  A magnetic field representation that applies to all magnetic fields allows features of magnetic evolution in toroidal and natural plasmas to be compared.  This comparison illustrates why plasmoid models of magnetic field evolution are not robust solutions to Maxwell's equations.  Despite five-hundred papers on plasmoid models, they provide a problematic description of fast magnetic reconnection.

\end{abstract}

\date{\today} 
\maketitle



Magnetic fields that are embedded in many natural and laboratory plasmas evolve in the near-ideal limit.  Although the breaking of magnetic field line connections is mathematically impossible if the evolution is truly ideal, magnetic reconnection is commonly observed to proceed at a rate determined by Alfv\'enic, not resistive physics \cite{Liu:2017}.  A standard explanation uses two-dimensional plasmoid models.  See the reviews \cite{Zweibel:review,Loureiro:2016}.  Nevertheless, actual physical space is three dimensional.   Reconnection in three dimensions differs fundamentally from that in two either due to turbulence \cite{Turbulent reconnection} or to the characteristic exponentially increasing sensitivity to non-ideal effects \cite{Boozer:Prevalent2018,Boozer:ideal B 2019}.   

In three dimensions, but not in two, a magnetic field that is evolving ideally, $\partial\vec{B}/\partial t=\vec{\nabla}\times (\vec{u}\times\vec{B})$, will become exponentially more sensitive to non-ideal effects as time advances \cite{Boozer:Prevalent2018,Boozer:ideal B 2019} unless the velocity \cite{Newcomb} of the magnetic field lines $\vec{u}(\vec{x},t)$ has a truly exceptional form.  When non-ideal effects are sufficiently small, systems naturally evolve into a state at which reconnection proceeds Alfv\'enically, not resistively \cite{Boozer:Prevalent2018,Boozer:ideal B 2019}.  

An additional problem in plasmoid models is that the location of the reconnection is a line. Viewed from a three-dimensional prospective, plasmoid models have two perfect symmetries: one orthogonal to the reconnection plane and another along a line in that plane, the line along which the plasmoids will form.  Three examples of the approximately five-hundred papers on plasmoid models are \cite{Loureiro:2007,Liu:2017,Huang:2017}.  In toroidal plasmas, it is known that the location and the nature of the reconnection is very sensitive to the perfection of both symmetries.  What will be shown is that plasmoid theory is not robust, which means it can be fundamentally changed by arbitrarily small magnetic perturbations.  This makes its application to natural plasmas problematic.

Maxwell's equations determine the evolution of magnetic fields whether embedded in toroidal laboratory or in natural plasmas.  A magnetic field representation that applies to all magnetic fields allows features of magnetic evolution in toroidal and natural plasmas to be compared.  This comparison illustrates why plasmoid models of magnetic field evolution in space are not robust.

A magnetic field in three-dimensional space always has the representation
  \begin{equation}
  2\pi\vec{B} = \vec{\nabla}\psi_t \times \vec{\nabla}\theta +\vec{\nabla}\varphi \times \vec{\nabla} \psi_p. \label{B-rep}
  \end{equation} 
 This representation is well known in toroidal plasma physics, where the interpretations are that $\theta$ is a poloidal angle, $\varphi$ is a toroidal angle,  $\psi_t$ is the toroidal magnetic flux enclosed by a constant-$\psi_t$ surface, and $\psi_p$ is the poloidal magnetic flux outside of a constant-$\psi_p$ surface \cite{Boozer:RMP}.
 
 The only requirements for the validity of Equation (\ref{B-rep}) are that $\vec{\nabla}\cdot\vec{B}=0$ everywhere and that  well behaved spatial coordinates $(r,\theta,\varphi)$ exist.  It is irrelevant whether the coordinates $\theta$ and $\varphi$ are angle like.   The proof is simple and can be skipped.  The Poincar\'e lemma states that a well behaved vector potential $\vec{A}(\vec{x},t)$ exists such that $\vec{B}=\vec{\nabla}\times\vec{A}$ when $\vec{\nabla}\cdot\vec{B}=0$ everywhere.  The vector potential for a magnetic field can be written, as can any vector in three-dimensional space, as $\vec{A} = A_r \vec{\nabla}r + A_\theta \vec{\nabla}\theta + A_{\varphi}\vec{\nabla}\varphi$, so
  \begin{eqnarray}
\vec{A} = \vec{\nabla}g + \frac{\psi_t }{2\pi}\vec{\nabla}\theta - \frac{\psi_p }{2\pi}\vec{\nabla}\varphi, \label{A-rep} 
\end{eqnarray}
where $\partial g/\partial r \equiv A_r$, $\psi_t /2\pi = A_\theta - \partial g/\partial \theta,$ and $\psi_p/2\pi \equiv  - A_\varphi + \partial g/\partial \varphi.$  The curl of Equation (\ref{A-rep}) yields Equation (\ref{B-rep}).

The $\vec{B}\cdot\vec{\nabla}\varphi$ component of the magnetic field is called the guide field in reconnection theory, and the magnetic field given by $\psi_p$ is called the reconnecting field.   The plasmoids form where $\vec{\nabla}\psi_p(r,\theta,t)=0$.  Symmetry in $\varphi$ implies neither $\psi_p$ nor $\psi_t$ depends on $\varphi$. 

The toroidal field, $\vec{B}\cdot\vec{\nabla}\varphi$, is non-zero everywhere in tokamak and stellarator plasmas.  In standard plasmoid theory, the guide field may be either zero or not at the place where reconnection will take place.  Nevertheless, the only robust state that is independent of $\varphi$ has a non-zero $\vec{B}\cdot\vec{\nabla}\varphi$ where $\vec{\nabla}\psi_p(r,\theta)=0$.   The condition $\vec{\nabla}\psi_p(r,\theta,t)=0$ is equivalent to $\partial\psi_p/\partial r=0$ and $\partial\psi_p/\partial \theta=0$.  That is, the function $\psi_p(r,\theta)$ must satisfy two conditions.  Solution points $(r,\theta)$ of two equations with two unknowns are generally well separated.  When solution points are not well separated, an arbitrarily small perturbation $\tilde{\psi}_p(r,\theta)$ to $\psi_p$ can make them so, which is what is meant by a solution is not being robust.  With symmetry in $\varphi$, the condition that $\vec{B}\cdot\vec{\nabla}\varphi=0$ is that $\partial \psi_t(r,\theta)/\partial r=0$.  Since the only robust solutions for locations where the reconnecting field vanishes are well separated $(r,\theta)$ points, there is no robust solution to the condition that $\partial \psi_t(r,\theta)/\partial r=0$ at the same $(r,\theta)$ points.   

Where $\vec{B}\cdot\vec{\nabla}\varphi\neq0$, positions in ordinary Cartesian coordinates can be given by the function $\vec{x}(\psi_t,\theta,\varphi,t)$ since the Jacobian $\mathcal{J}_c = 1/(\vec{\nabla}\psi_t \times \vec{\nabla}\theta)\cdot\vec{\nabla}\varphi$ is well behaved; $1/\mathcal{J}_c=2\pi \vec{B}\cdot\vec{\nabla}\varphi$.   The $(\psi_t,\theta,\varphi,t)$ coordinates are called canonical because in these coordinates magnetic field lines are given by Hamilton's equations \cite{Boozer:RMP}.  Along a field line $d\psi_t/d\varphi =\vec{B}\cdot\vec{\nabla}\psi_t / \vec{B}\cdot\vec{\nabla}\varphi$.  Equation (\ref{B-rep}) implies $d\psi_t/d\varphi =-\partial\psi_p/\partial \theta$ and analogously $d\theta/d\varphi = \partial\psi_p/\partial \psi_t$.

In regions in which $\vec{B}\cdot\vec{\nabla}\varphi\neq0$, Faraday's law implies that the electric field can always be written as
\begin{eqnarray} 
&&\vec{E} +\vec{u}_c\times\vec{B} = \left(\frac{\partial\psi_p}{\partial t}\right)_c \frac{\vec{\nabla}\varphi}{2\pi} - \vec{\nabla}\Phi_c,  \label{psi-p-ev} 
\end{eqnarray}
where $\vec{u}_c \equiv \left(\partial \vec{x}/\partial t\right)_c$ is the velocity of the canonical coordinates through space.  The subscript ``c" implies a dependence on canonical coordinates.  Equation (\ref{psi-p-ev}) is easily derived using Equation (A19) from the appendix of \cite{Boozer:RMP}. 

The function $\vec{x}(\psi_t,\theta,\varphi,t)$ defines a homotopy in space as time $t$ evolves, which ensures magnetic field lines cannot change their topology unless $\psi_p$ changes.  Therefore, Equation (\ref{psi-p-ev}) implies magnetic field lines will not change their topology if a function $\Phi_c(\vec{x},t)$ exists, such that 
\begin{equation} 
\vec{B}\cdot\vec{\nabla}\Phi_c=-\vec{E}\cdot\vec{B}. \label{Phi_c eq}
\end{equation}

Dotting Equation (\ref{psi-p-ev}) with $\vec{B}$ and integrating over a volume, $d^3x=\mathcal{J}_c d\psi_t d\theta d\varphi$, one finds that
\begin{equation}
\int \vec{E}\cdot\vec{B}d^3x=\int \left(\frac{\partial\psi_p}{\partial t}\right)_c\frac{d\psi_t d\theta d\varphi}{(2\pi)^2} -\oint \Phi_c \vec{B}\cdot d\vec{a}.
\end{equation}
This equation implies the magnetic helicity, which is proportional to $\int \psi_p d\psi_t d\theta d\varphi$ between two magnetic surfaces (surfaces on which $\vec{B}\cdot d\vec{a}=0$), can change only on a resistive or other non-ideal time scale given by $\int \vec{E}\cdot\vec{B}d^3x$.  Properties of the magnetic field that depend on $\vec{x}(\psi_t,\theta,\varphi,t)$, such as the spatial distribution of the plasma current, can change on an ideal time scale limited by the Alfv\'en speed.  Such well separated time scales are seen in tokamak experiments \cite{Boozer:pivotal,de Vries:2016}.

Axisymmetric tokamaks are symmetric in $\varphi$, which is the only symmetry possible in spatially bounded near-Maxwellian plasmas.  An important feature of $\varphi$ symmetry, whether $\varphi$ is a toroidal angle or not, is that
\begin{equation} 
\vec{B}\cdot\vec{\nabla}\psi_p=0.
\end{equation}
Constant-$\psi_p$ surfaces are magnetic surfaces.  This is an implication of Equation (\ref{B-rep}).  When $\psi_p$ and $\psi_t$ depend on only the $r$ and $\theta$ spatial coordinates, $2\pi\vec{B}\cdot\vec{\nabla}\psi_p = (\partial\psi_t/\partial r) (\vec{\nabla}r\times\vec{\nabla}\theta)\cdot\vec{\nabla}\psi_p=0$. 

Magnetic reconnection that is fast compared to the time scale given by $\int \vec{E}\cdot\vec{B}d^3x$ is not robustly possible in systems with $\varphi$ symmetry, such as those in plasmoid theory.  The integral $\int\psi_pd\psi_td\theta d \varphi=\int\psi_p(\partial\psi_t/\partial r) dr d\theta d\varphi$ between any two constant-$\psi_p$ contours can only change on the intrinsically slow non-ideal time scale determined by $\int \vec{E}\cdot\vec{B}d^3x$, not on the Alfv\'enic time that can characterize magnetic relaxations---particularly after a magnetic reconnection.  As shown in \cite{Boozer:Prevalent2018,Boozer:ideal B 2019} a fast magnetic reconnection can occur over a large volume at an Alfv\'enic rate with $\int \vec{E}\cdot\vec{B}d^3x$ arbitrarily small when $\varphi$-symmetry is broken.   In standard plasmoid theory, $\int \vec{E}\cdot\vec{B}d^3x=\int\eta \vec{j}\cdot\vec{B} d^3x$ becomes sufficiently large to give Alfv\'enic reconnection.  No natural separation exists between the helicity conserving and the resistive time scales that is seen in tokamak disruptions \cite{Boozer:pivotal,de Vries:2016}.  It is also true \cite{Boozer:ideal B 2019} that in three-dimensions, the current density increases exponentially slower than the sensitivity of reconnection to  even a fixed $\int \vec{E}\cdot\vec{B}d^3x$.  

The saddle points of $\psi_p(r,\theta,t)$ are special.  At these points, which are also called $X$ points, $\partial\psi_p/\partial r=0$ and $\partial\psi_p/\partial \theta=0$ but without $\psi_p$ having a maximum or a minimum.  Field lines that pass through an $X$ point exponentiate apart and define the separatrices of islands on the constant-$\psi_p$ surfaces.   As shown, in systems that are sufficiently robust to arise naturally, the $X$-points of $\psi_p$ can be assumed to be well separated and generally not on the same constant-$r$ curve.  

Magnetic reconnection has special properties at $X$-points  when the magnetic field is perturbed while preserving $\varphi$ symmetry \cite{Grad:1975,Birn:2007}.  This special type of reconnection arises in a tokamak with a divertor in an axisymmetric vertical instability \cite{Pfefferle:2018} but, nevertheless, has a minor role in the tokamak literature.  Despite its minor role in the tokamak literature, this special type of reconnection is the basis of the plasmoid theory.  Indeed, the standard situation imagined in plasmoid theory, for example \cite{Loureiro:2007,Liu:2017,Huang:2017}, is even more special, an initial state in which perfect symmetry also exists in $\theta$ on the line along which reconnection will occur, even when the overall system does not have that symmetry. 

Plasmoids form along the line where $\vec{\nabla}\psi_p$ vanishes.  Whether $\vec{\nabla}\psi_p$ vanishes along a line or at well separated points, which is the only robust solution, makes a major change in the theory of reconnection.  Consider axisymmetric tokamaks, $\vec{\nabla}\psi_p$ vanishes for all $\theta$ on a magnetic surface on which the twist of the magnetic field lines, or rotational transform, $\iota\equiv \partial \psi_p(\psi_t,t)/\partial\psi_t$ changes sign.   The $\psi_t$ that appears in $\psi_p(\psi_t,t)$ is the toroidal magnetic flux enclosed by a constant $\psi_p$ surface.  Tokamak equilibria break  poloidal symmetry by terms of order $a/R$, the minor radius divided by the major radius, and a scalar pressure equilibrium in which $\iota$ changes sign across a magnetic surface is not possible \cite{Hammett:2004}.  In cylindrical plasmas such equilibria are possible due to the perfect symmetry in $\theta$ and $\varphi$.  Equilibria in which $\iota$ has both signs are possible in an axisymmetric tokamak, when $\vec{\nabla}\psi_p$ vanishes only at points \cite{Hammett:2004,Martynov:2003,Wang: 2004}.  Such solutions have axisymmetric magnetic-island-like structures.  That is, the poloidal flux has the functional form $\psi_p(r,\theta,t)$ and cannot be expressed as a global single-valued function of the toroidal flux and time, although locally it can be.  Despite the existence of tokamak equilibria in which $\iota$ has both signs, experiments on JET were not able to access such equilibria, and numerical simulation implied $\iota$ reversal was impeded by rapid magnetic reconnection \cite{Stratton:2002}.  What has apparently not been done are careful studies of the resistive evolution of axisymmetric toroidal equilibria in which $\iota$ reverses across the plasma with the two signs of $\iota$ separated by an axisymmetric island chain.

The standard studies using plasmoids have reconnection occurring on a surface that has a symmetry the system itself does not have.  Based on the results of Hammett et al \cite{Hammett:2004}, it appears that such equilibria do not exist.  It seems unlikely that plasmas without scalar pressure equilibria would evolve slowly into a state of fast magnetic reconnection with the frequency that such events are seen.  

The most obvious difference between pictures of magnetic reconnection in toroidal plasmas and of plasmoid reconnection is the location of islands at well separated rational surfaces in tori, which means $\iota$ is the ratio of two integers, and at a single surface in plasmoid reconnection.  In tokamaks, islands are often observed to evolve slowly, and fast magnetic reconnection only ensues when islands on distinct surfaces have widths comparable to their separations.  This qualitative behavior was seen on JET \cite{de Vries:2016}.  A growth of non-axisymmetric perturbations associated with magnetic islands was observed to occur over approximately a half second followed by a spike in the plasma current and a large drop in the internal inductance on a time scale about a thousand times shorter.  The long time scale can be interpreted as resistive and the thousand times faster time scale has the features of a helicity-conserving breakup of the magnetic surfaces through a fast magnetic reconnection \cite{Boozer:pivotal}. 

The separation of islands is easily explained in toroidal plasmas.   Equation (\ref{psi-p-ev}) implies an ideal evolution $(\partial\psi_p/\partial t)_c=0$ is not possible when the magnetic field lines on a constant-$\psi_p$ surface close on themselves unless the loop voltage $V_\ell\equiv \oint E_{||}d\ell/m=0$.  The differential distance along the line is $d\ell$, and field lines on rational surfaces close on themselves after $m$ toroidal and $n$ poloidal circuits with $\iota=n/m$.  A rational surface will split to form an island chain unless $V_\ell$ has exactly the same value on every line in the surface.  The evolution is fundamentally different when $\iota$ is an irrational number and a magnetic field line neither closes on itself nor leaves a spatially-bounded constant-$\psi_p$ surface.  Then, each line comes arbitrarily close to every other line on the surface, and $\psi_p$ changes at exactly the same rate everywhere on the irrational surface, $V_\ell\equiv \int E_{||}d\ell/m$ as $m\rightarrow\infty$.

The situation is more problematic in plasmoid models in which $\theta$ and $\varphi$ are not angles.  The issue with $\varphi$ periodicity can be appreciated by letting the major radius of a torus become ever larger and considering the effect of a finite Alfv\'en speed, $V_A$.  When $2\pi R/V_A$ is much longer that the time scale over which a non-$\varphi$ symmetric perturbation has its effect, then different parts of the torus do not interact magnetically.  Each part responds with a locally defined parallel current density.  The equation $\vec{\nabla}\cdot\vec{j}=0$ requires $\vec{B}\cdot\vec{\nabla}(j_{||}/B)=\vec{B}\cdot\vec{\nabla}\times(\vec{f}_L/B^2)$, but this can be satisfied by equating the Lorentz force, $\vec{f}_L\equiv\vec{j}\times\vec{B}$, to the inertial force of an Alfv\'en wave.   The evolution of a plasma into a rapidly reconnecting state is sensitive to symmetry breaking over the distance scale that an Alfv\'en wave can traverse during an evolution time.

A fundamental question in plasmoid models is why the current density becomes large where a component of the magnetic field passes through zero.  An example illustrates a problem that arises when no direction perpendicular to a line along which the plasmoids form has perfect symmetry.  Consider the force-free field, $\vec{B}/B_0=\sin(x/a)\hat{y} + \cos(x/a)\hat{z}$ with $B_0$ and $a$ constants.   This field reverses its direction in the $x$-$y$ plane at $x=0$.  Rotate the point of view by an angle $\alpha$ about the $\hat{x}$ axis and define new Cartesian basis vectors by $\hat{x}=\hat{X}$, $\hat{y} =\hat{Y} \cos\alpha + \hat{Z}\sin\alpha$, and $\hat{z} =\hat{Z} \cos\alpha - \hat{Y}\sin\alpha$.  Then, $\vec{B}/B_0=\sin(x/a-\alpha)\hat{Y} + \cos(x/a-\alpha)\hat{Z}$, which vanishes in the $X$-$Y$ plane at $x/a=\alpha$.  In a sheared magnetic field, the field-reversal location in a plane can be chosen to be at any desired value of $x$ by picking the angle of observation to be $\alpha=x/a$.   

The plasma response in fast magnetic reconnection, from current sheets to hall currents, is too complicated to realistically compute.  What can be done is to focus on the features of fast magnetic reconnection that are robust---the features that persist despite uncertainties in the physics model and without exact symmetries.  Three dimensional calculations have this robustness and explain why fast magnetic reconnection should naturally arise in near-ideal, evolving plasmas \cite{Boozer:Prevalent2018,Boozer:ideal B 2019}; plasmoid models do not.

\section*{Acknowledgements}

This material is based upon work supported by the U.S. Department of Energy, Office of Science, Office of Fusion Energy Sciences under Award Numbers DE-FG02-95ER54333, DE-FG02-03ER54696, DE-SC0018424, and DE-SC0016347.


\end{document}